%% file: PEEK_arxiv.tex
\documentclass[sigconf, screen]{acmart}
\usepackage{tikz}

\usepackage{booktabs}
\usepackage{url}
\usepackage{tabularx}
\usepackage{tcolorbox}
\usepackage{makecell}
\usepackage{colortbl}
\usepackage{multirow}
\AtBeginDocument{%
  }

\setcopyright{none} 
\copyrightyear{2026}
\acmYear{2026}
\acmDOI{XXXXXXX.XXXXXXX}

\acmConference[MICRO 2026]{The 58th IEEE/ACM International Symposium on Microarchitecture}{October 31--November 04, 2026}{Athens, Greece}

\acmISBN{978-X-XXXX-XXXX-X/XX/XX}

\setcopyright{none}
\newcommand{\eg}{e.g.}
\newcommand{\ie}{i.e.}
\newcommand{\etc}{etc.}
\newcommand{\defn}[1]{\textbf{\textit{#1}}}

\newcommand{\ignore}[1]{}

\makeatletter
\renewcommand{\@makecaption}[2]{%
  \vspace{3pt}   
  \sbox\@tempboxa{\textbf{#1: #2}}%
  \ifdim \wd\@tempboxa >\hsize
    \textbf{#1: #2}\par
  \else
    \global \@minipagefalse
    \hb@xt@\hsize{\hfil\box\@tempboxa\hfil}%
  \fi
  \vspace{5pt}   
}
\makeatother

\newrobustcmd*\blacka[1]{\tikz[baseline=(char.base)]{
            \node[shape=circle,draw,inner sep=1pt,fill,text=white,minimum size=1em] (char) {\textsf{\small a}};}}

\newrobustcmd*\blackb[1]{\tikz[baseline=(char.base)]{
            \node[shape=circle,draw,inner sep=1pt,fill,text=white,minimum size=1em] (char) {\textsf{\small b}};}}

\newrobustcmd*\blackc[1]{\tikz[baseline=(char.base)]{
            \node[shape=circle,draw,inner sep=1pt,fill,text=white,minimum size=1em] (char) {\textsf{\small c}};}}

\newrobustcmd*\blackd[1]{\tikz[baseline=(char.base)]{
            \node[shape=circle,draw,inner sep=1pt,fill,text=white,minimum size=1em] (char) {\textsf{\small d}};}}

\newrobustcmd*\blacke[1]{\tikz[baseline=(char.base)]{
            \node[shape=circle,draw,inner sep=1pt,fill,text=white,minimum size=1em] (char) {\textsf{\small e}};}}

\newrobustcmd*\blackf[1]{\tikz[baseline=(char.base)]{
            \node[shape=circle,draw,inner sep=1pt,fill,text=white,minimum size=1em] (char) {\textsf{\small f}};}}

\newrobustcmd*\blackg[1]{\tikz[baseline=(char.base)]{
            \node[shape=circle,draw,inner sep=1pt,fill,text=white,minimum size=1em] (char) {\textsf{\small g}};}}
\begin{document}

\title{PEEK: Heterogeneous Parallelism for Privileged Error Detection in Safety-Critical Processors}




\author{\small Tinglue Wang\textsuperscript{2,3}, Zhenghui Guo\textsuperscript{2,3}, Bing Guo\textsuperscript{2,3}, Jiapeng Guan\textsuperscript{2,3}, Renshuang Jiang\textsuperscript{4}, Jie Zhang\textsuperscript{2,3}, Jing Li\textsuperscript{5}, Xin Si\textsuperscript{2,3}, Sam Ainsworth\textsuperscript{6}, Zhe Jiang\textsuperscript{2,1,3}}

\affiliation{\institution{\small \textsuperscript{2}Southeast University, China; \textsuperscript{3}National Center of Technology Innovation for EDA, China; \textsuperscript{4}National University of Defense Technology, China; \textsuperscript{5}New Jersey Institute of Technology, US; \textsuperscript{6}University of Edinburgh, UK.}\country{China}}

\thanks{\textsuperscript{1}Corresponding author. Emails: 101013615@seu.edu.cn.}

\renewcommand{\shortauthors}{Wang et al.}

\begin{abstract}

  Heterogeneous parallel error detection architecture has been widely studied for safeguarding OoO superscalar processors in safety-critical systems, as it achieves significantly lower hardware overhead compared to traditional LockStep, by exploiting the parallelism that exists in a secondary execution.
However, previous works do not cover the protection of privileged-mode execution, impeding their effectiveness in real-world deployment. Moreover, naive extension to privileged-mode can cause a litany of issues, from abysmal performance due to high synchronization costs, to full deadlocks.

Here, we present PEEK, the first privileged parallel error detection architecture.
Based on a deep analysis of privileged execution, we redesign the verification pipeline, addressing all the bottlenecks and bugs identified in privileged-mode protection.
Evaluated using various metrics on an RTL-level full system running Linux, PEEK achieves full-privilege protection on Linux with negligible performance slowdown and affordable hardware overhead. PEEK has been taped out using a 28nm process, and its source is available at https://anonymous.4open.science/r/PEEK-3000.

\end{abstract}


\keywords{Hardware Fault Tolerance, Safety-Critical Systems, Full System}

\maketitle

\input{partition/intro}

\input{partition/background}

\input{partition/motivation}
\input{partition/overview}
\input{partition/micro}
\input{partition/deadlock}
\input{partition/evaluation}

\input{partition/conclusion}

\newpage


\bibliographystyle{ACM-Reference-Format}
\bibliography{ref}

\end{document}

%% file: partition/intro.tex
\section{Introduction}
Safety-critical systems, from automated vehicles to industrial controllers, demand extremely reliable processors that can withstand harsh conditions~\cite{wang2023safety,sim2020condition, perez2015condition, athavale2019condition}, such as extreme temperatures and cosmic radiation.
Meanwhile, continued downscaling of semiconductor technology, with increased transistor density and lower voltages, has led to a rising incidence of both transient and permanent hardware faults in processors~\cite{hernandez2015timely, romanescu2008core, reinhardt2000transient, austin1999diva,coresdontcount,sdc_at_scale,dixit2022detectingsilentdatacorruptions}.
To meet stringent safety requirements, today's safety-critical systems often employ hardware-based \defn{LockStep} mechanisms~\cite{lockstep2015,arm-triple-lock,iturbe2019arm, marcinek2023lockstep}, which duplicate the processor core, execute the same program, and compare outputs of core’s pins at each clock cycle. 
While LockStep provides strong assurance by replaying everything on a synchronized secondary core and verifying execution at the signal level, it incurs substantial costs, more than twice the area and power of a single core.
These overheads become more pronounced when safety-critical systems increasingly need \defn{Out-of-Order} (\defn{OoO}) superscalar processors~\cite{yarza2022HPC, knight2002HPC,alcaide2023HPC, reinhardt2019HPC} to meet growing demands for computational throughput and instruction-level parallelism.


\begin{figure}[t]
  \centering
  \includegraphics[width=1\columnwidth]{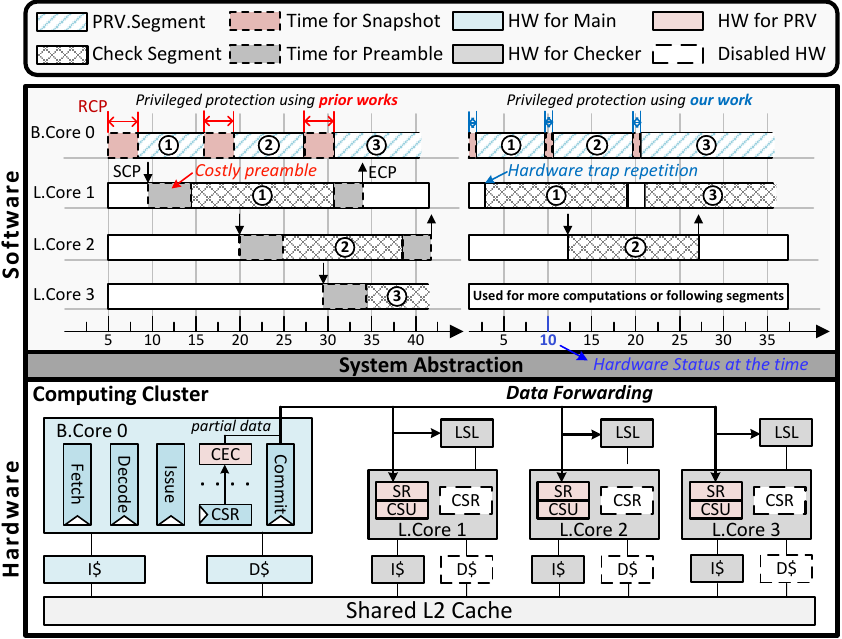}
  \caption{PEEK enables HPED to cover privileged execution (new units in pink;  \textit{RCP: Register Checkpoint; S/ECP: Start/End RCP; LSL: Load-Store Log; CEC: CSRs Extraction Controller; SR: Shadow Registers; CSU: Context Switching Unit}). 
  PEEK uses partial snapshots that duplicate only CSRs actually used, rather than the hundreds of CSRs that are not used, reducing the back-pressure. Hardware trap repetition allows efficient state transitions while only duplicating real work, avoiding costly preambles. 
  In this example, two little cores are sufficient to keep up with the big core in our work, whereas prior works~\cite{jiang2025meek, ainsworth2018parallel} require a third little core.
  }
  \label{top}
\end{figure}

\textbf{Heterogeneous Parallel Error Detection (HPED)} architecture~\cite{ainsworth2018parallel, ainsworth2019paramedic, ainsworth2021paradox, liao2025paraverser} has emerged as a promising alternative for safeguarding OoO superscalar processors.
HPED divides a software program running on a high-performance OoO core (big core) into multiple discrete segments using \defn{Register Checkpoints} (\defn{RCPs}) and re-executes these segments on a set of smaller, energy-efficient cores (little cores) for verification (Figure~\ref{top}: software), later reconstructing the full run using induction. This approach achieves error detection with significantly lower hardware overhead compared to LockStep.
While prior work has demonstrated the conceptual feasibility of HPED, \eg, reporting 1.02$\times$ slowdown and 24\% area overhead~\cite{ainsworth2018parallel}, these results were obtained using highly abstracted syscall-emulation simulations~\cite{ainsworth2018parallel, ainsworth2019paramedic, ainsworth2021paradox, liao2025paraverser} or minimal prototypes with idealistic assumptions~\cite{jiang2025meek}. Moreover, existing designs lack full Operating System (OS) support and can only run the workloads in unprivileged mode, without protection of privileged execution.

\smallskip\noindent
\textbf{Challenges.}
In our effort to tape out a real chip using this architecture with full Linux support, we encountered critical challenges related to privileged execution. Although privileged execution typically constitutes a small fraction of execution time, absence of protection creates a substantial vulnerability in system safety. 
We initially attempted to apply the same methodology~\cite{jiang2025meek} for privileged protection, but found that naively replaying privileged execution led to severe performance degradation and even system-level deadlocks.
Deeper analysis revealed that privileged execution exhibits fundamentally different characteristics from user-mode code: it frequently interacts with hardware state through \defn{Control and Status Register} (\defn{CSR}) operations, and handles asynchronous events like interrupts. 
As a result, synchronizing only generic architectural registers is insufficient to faithfully replicate privileged state. On the other hand, capturing the complete privileged context drastically expands the snapshot size, which imposes substantial back-pressure in realistic transmission units and ultimately leads to severe slowdown (Figure~\ref{top}: B.Core).
Additionally, using native trap handlers for replaying privileged execution incurs high computational overhead on little cores, as they must spend significant time on preparing privileged-mode entry and exit, instead of verification (Figure~\ref{top}: L.Core).
Finally, contention between the big core and little cores over shared OS locks~\cite{jain2020contention,wang2020contention,zhuravlev2010contention}  that cannot be load-store logged~\cite{ainsworth2018parallel}, e.g., page-fault handling, complicates both OS-level resource management and verification scheduling, which can lead to system deadlocks (detailed in Section~\ref{motiv-deadlock}).

\smallskip\noindent
\textbf{Contributions.}
By systematically addressing the identified bottlenecks, we propose \defn{PEEK} (\defn{Privileged-Execution Error-Keeper}), the first \textit{privileged} parallel error detection architecture.
Through a deep analysis of privileged-mode behavior, PEEK efficiently replicates the big core’s privileged state with minimal bandwidth and area overhead. 
To support lightweight privileged replay on little cores, we design a mechanism mimicking standard hardware event handling, thereby avoiding unnecessary computational costs. Additionally, PEEK leverages existing modules to proactively resolve resource contention, ensuring deadlock-free execution in privileged mode.
We implement PEEK on an AMD Alveo U280 FPGA, booting a full Linux kernel (v6.7) to validate its practical feasibility. Experimental results show that PEEK achieves full privilege-mode coverage with only a 1.02$\times$ performance slowdown,  microsecond-level detection latency, and area overhead of 1.8\% on the big core and 3.8\% on the little cores.
We have taped out PEEK with a smaller configuration (detailed in Section~\ref{tapeout}) and publicly released PEEK’s source code, at https://anonymous.4open.science/r/PEEK-3000.

%% file: partition/background.tex
\section{Background}

\subsection{Detection is Key}
To mitigate errors caused by hardware faults, protection mechanisms, from error detection codes~\cite{haring2021code, hillier2019code} to fault-tolerant architectures~\cite{pasham2020architecture,ahangari2023architecture}, have been developed.
Among them, error detection serves as the foundation~\cite{le2018detection,rausand2014detection,singh2025detection, gizopoulos2011detection, raychowdhury2011detection}: 
once an error is identified, the system can transition to a safe state and initiate corrective actions (e.g., fault isolation~\cite{de2019isolating,hale2019isolating} or state rollback~\cite{arm2016rollback,gajre2025rollback}) to prevent escalation into hazardous failures.
As mandated by industrial safety standards, e.g., ISO 26262~\cite{iso201126262} for automotive and DO-178C~\cite{rierson2017developing} for avionics, mitigation procedures must complete within a Fault Tolerance Time Interval (on the order of milliseconds~\cite{jiang2018bluevisor}).

Software-based solutions (e.g., instruction redundancy~\cite{oh2002instred,he2024instred,didehban2016nzdc,parallaft} and software scanners~\cite{batcher1999scanner,bernardi2015sscanner,dixit2022detectingsilentdatacorruptions,serebryany2021silifuzzfuzzingcpusproxy}) suffer from limited fault coverage, lack of multithreading support and substantial performance degradation, making them inadequate for stringent safety standards.
In contrast, LockStep utilizes full hardware redundancy to provide utmost safety protection, and hence is deployed in relatively smaller (in-order, microcontroller-sized) commercial processors, such as Renesas's RH850~\cite{RenesasRH850} and ARM's Cortex-R52~\cite{Cortex-R5}.
However, duplicating OoO superscalar cores is prohibitively expensive~\cite{de2017higharea, mach2024higharea,kempf2021adhigharea,barbirotta2024higharea}, as their complex microarchitectures increase area and power by orders of magnitude.
Moreover, the cycle-by-cycle, pin-level comparison required by LockStep imposes long, high-fanout interconnects that complicate physical design and timing closure, reduce achievable clock frequency, and destabilize silicon bring-up, undermining the advantages of OoO processors.

\begin{figure}[t]
 \centering
 \includegraphics[width=1\columnwidth]{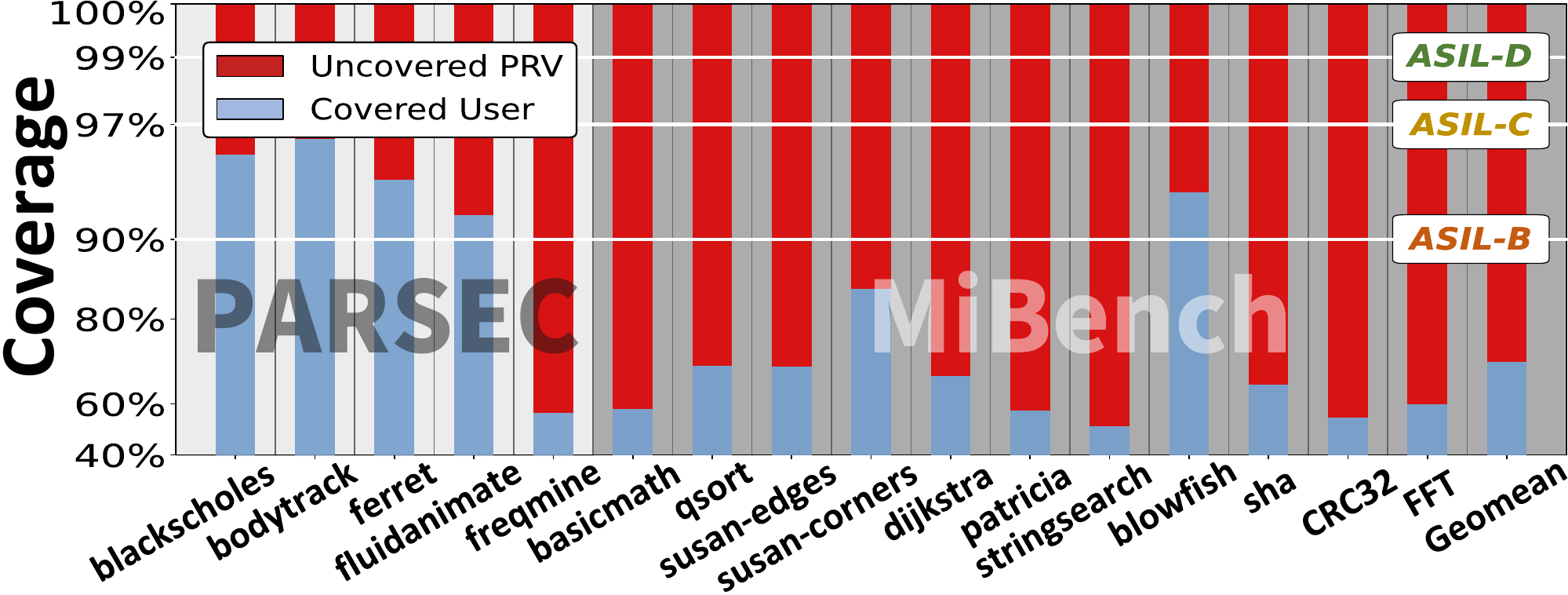}
 \caption{Measured fault coverage and ASIL region. }
 \label{asil}
\end{figure}

\subsection{Heterogeneous Parallel Error Detection}
In contrast to the synchronous re-execution of LockStep, HPED~\cite{ainsworth2018parallel} achieves parallel re-execution by dividing the program running on a big core into checking segments using register checkpoints (RCPs), \ie, architectural state captured periodically, and replaying them asynchronously and overlapped on multiple slower little cores (Figure~\ref{top}: software).
Little cores redirect their memory accesses from the D-Cache to the \defn{Load-Store Log} (\defn{LSL}) to re-read and check memory accesses as they were at the time of big-core execution (Figure~\ref{top}: hardware). Before executing a checking segment, a little core initializes its architectural state to the segment's \defn{Start RCP} (\defn{SCP}) and compares its final state to the \defn{End RCP} (\defn{ECP}) upon completion. 
If all little cores are busy processing segments, the big core must stall until a little core becomes available.
The program is measured correct if all segments' ECPs and memory data match the original execution, and ECPs of segment $N$ match SCPs of segment $N+1$, to provide a proof-by-induction of equivalence.

The recent MEEK~\cite{jiang2025meek} presents the first full-RTL HPED design with a complete microarchitecture implementation.
MEEK includes a non-intrusive Data Extraction Unit in the big core for data collection and a Scheduling Engine to orchestrate little core resources. 
In little cores, a Mode Switch Unit and a LSL enable the replay of checking segments. 
It also features a Forwarding Fabric to alleviate transmission bottlenecks. 
However, even with its full-RTL implementation, MEEK still remains a user-mode prototype and does not provide privileged protection: any privileged code running on big core is ignored by little cores, restarting checks on return to user.

%% file: partition/motivation.tex
\section{Motivation}
\label{write-motivation}

\subsection{Privileged Gaps Threaten System Safety}
\label{motiv gap}
Prior HPED architectures~\cite{ainsworth2018parallel, liao2025paraverser, jiang2025meek} focus solely on user-mode execution, with implementations and evaluations confined to user mode, using gem5 syscall-emulation simulations without a full OS~\cite{ainsworth2018parallel,liao2025paraverser}, or codesigning implementation parts into software processes and reusing OS functionality without checks~\cite{jiang2025meek}. This leaves a critical blind spot regarding their full-system safety protection.

To quantify this blind spot, we deploy a full Linux on an OoO superscalar processor and assess fault coverage across representative workloads (see Section~\ref{eval} for the full experimental setup): PARSEC~\cite{bienia2008parsec} for its general-purpose, user-mode-intensive workloads adopted in prior work~\cite{ainsworth2018parallel, ainsworth2019paramedic, ainsworth2021paradox, jiang2025meek}, and MiBench~\cite{guthaus2001mibench} for its safety-critical embedded workloads that stress privileged activities.
As shown in Figure~\ref{asil}, even user-mode-intensive PARSEC workloads fail to achieve the 97\% coverage mandated for ASIL-C (the second-highest ISO 26262 safety level).
The deficit is more severe in MiBench, where nearly all workloads fall below the 90\% coverage required for ASIL-B.
This means that without privileged-mode protection, HPED can only satisfy ASIL-A (the lowest ISO safety level), suitable only for non-critical functions (e.g., infotainment).

These findings motivated a deeper investigation into the role of privileged execution from a safety standpoint.
Our source-code analysis revealed that, beyond managing normal execution, the privileged execution (particularly the OS kernel) implements numerous critical safety-related functions. These include not only CPU-level but also system-wide mechanisms, e.g., ECC memory error handlers~\cite{iguchi2022ECC}, fault recovery~\cite{cojocar2019ECC}, watchdog timers that reset hung subsystems~\cite{dobel2012watches, kluge2018watchdog}, and resource isolation to prevent fault propagation~\cite{hunt2019isolation,thiyyakat2020isolation}.
Some mechanisms specific to safety-critical systems also rely on privileged operations.
For example, CiAO~\cite{hoffmann2014effectiveness} uses privileged logic to check the CRC32 integrity of all objects when entering kernel, for early memory corruption detection;
VirtuOS~\cite{nikolaev2013virtuos} employs kernel virtualization to isolate slices of kernel functionality into protected domains, enabling fast recovery without reboots.
Thus, leaving privileged execution unprotected poses a risk of catastrophic system failure.

\begin{figure}[t]
      \centering
      \includegraphics[width=1\columnwidth]{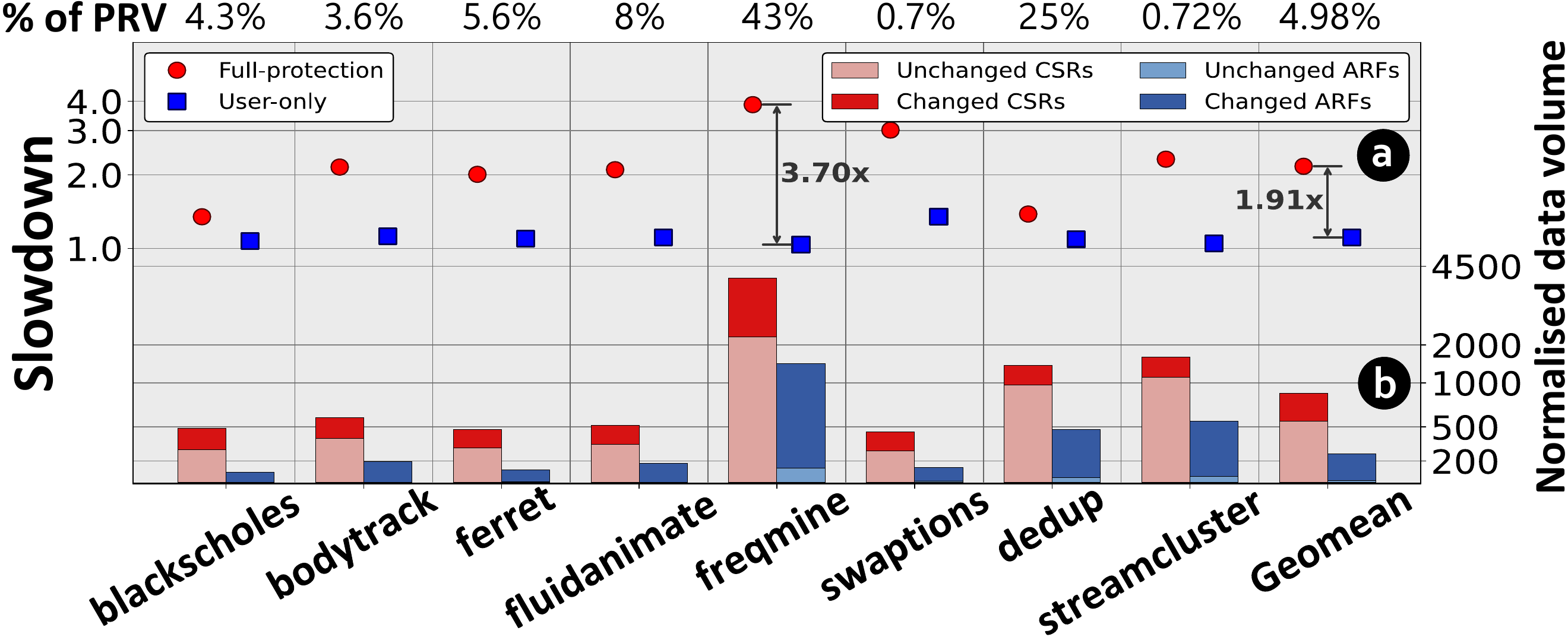}
      \caption{\blacka{a} Slowdown and \blackb{b} volume of different coverages. (Full protection also includes ARFs, which are omitted here for clarity; \% of PRV: proportion of privileged execution.)}
      \label{slowdown}
\end{figure}

\begin{tcolorbox}[
  colframe=black,
  boxrule=0.5pt,
  arc=1mm,
  left=1mm,
  right=1mm,
  top=0.5mm,
  bottom=0.5mm,
  boxsep=0.5mm
]

\textbf{\emph{Insight \#1}:} \textit{
Missing privileged protection not only causes insufficient fault coverage but also exposes the entire system to cascading failures. 
Error detection must be supported in privileged execution to ensure end-to-end system safety.
}
\end{tcolorbox}

\subsection{Snapshot Explosion Disrupts Verification }
\label{CSR snap}

Achieving privileged protection in HPED requires all 347 (up to 4,096 with customization~\cite{RISCV2025}) Control and Status Registers (CSRs) be captured in RCPs, in addition to the 64 architectural registers for user-only verification, leading to a 5.4$\times$ increase in snapshot size.  
This snapshot overhead has cascading impacts on processing efficiency across the entire verification pipeline from data extraction on big core, RCP transactions, status deployment (before checking), to status comparison and restoration (after checking) on little cores.

When applying the same methodology~\cite{ainsworth2019paramedic,jiang2025meek, ainsworth2021paradox} for full system protection, we observe a 1.91$\times$ geomean slowdown compared to user-only protection (Figure~\ref{slowdown} \blacka{a}), despite privileged mode accounting for only 4.98\% of total computation time. 
The slowdown peaks at 3.7$\times$ in freqmine, due to frequent \texttt{load}/\texttt{store} misalignments that trigger extensive privilege transitions. 
By analyzing snapshot volume across workloads, we find that slowdown is nearly proportional to the volume, meaning that checkpointing overheads dominate, motivating a reconsideration of the snapshot mechanism.

\begin{figure}[t]
      \centering
      \includegraphics[width=1\columnwidth]{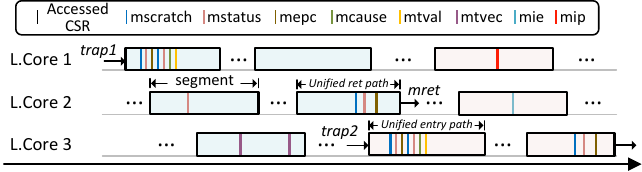}
      \caption{Trap-related CSR accesses in two timer-interrupt cases: except for the unified paths, \texttt{mstatus} and \texttt{mtvec} are accessed in trap1, while \texttt{mie} and \texttt{mip} are accessed in trap2.}
      \label{pattern}
\end{figure}

Unlike architectural registers, which frequently change across segments, a large subset of CSRs are seldom accessed during execution (Figure~\ref{slowdown} \blackb{b}).
For instance, over 60\% of CSRs are not accessed in any workload.
This suggests that capturing a partial snapshot, targeting only the actively accessed CSRs, could reduce inefficiencies.
However, Figure~\ref{pattern} shows a real-world example observed in our prototyping, revealing that the specific CSRs accessed can vary hugely even for the same trap cases.
Therefore, an online approach is needed to trace the behavior and selectively snapshot accessed CSRs. Crucially, it must maintain both behavioral consistency and complete verification, while minimizing snapshot size.

\begin{tcolorbox}[
  colframe=black,
  boxrule=0.5pt,
  arc=1mm,
  left=1mm,
  right=1mm,
  top=0.5mm,
  bottom=0.5mm,
  boxsep=0.5mm
]

    \textbf{\emph{Insight \#2}:} 
    \textit{Snapshotting CSRs as architectural registers degrades verification efficiency.
    A dynamic mechanism is required to selectively capture only accessed CSRs in RCPs.
    }

\end{tcolorbox}

\subsection{Native Traps Limit Privileged Replay}
\label{moti-OS}
    
\begin{figure}[t]
      \centering
      \includegraphics[width=\columnwidth]{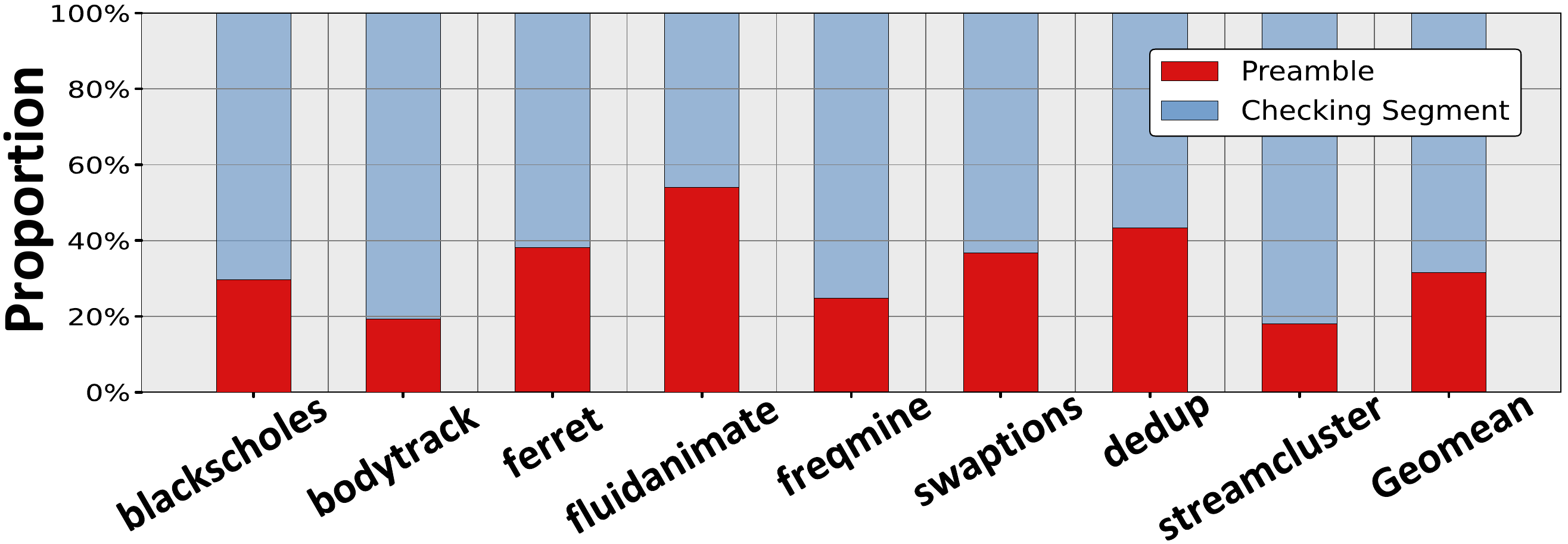}
      \caption{Comparison of execution time between trap-handler preambles and a checking segment in PARSEC.}
      \label{MSC comp}
    \end{figure}

Privileged-mode protection demands the little cores to replay fragmented checking segments under the same privilege mode as the big core. 
To do so, a little core must transition into the target mode, redirect itself to the segment's starting \defn{program counter} (\defn{PC}) before replaying, and revert after the check.
An intuitive approach is to leverage software-handled traps (\eg, ECALLs or interrupts) that are natively supported by the OS or firmware~\cite{cassano2022trap,ramsauer2022trap,voronetskiy2025trap}.
However, a trap invocation incurs a long preamble (saving the full context, decoding the trap cause via the vector table, branching into the handling routines, etc.), which consumes more than 2,000 cycles on each trigger, significantly reducing little cores' utilization and wasting verification capacity.
As quantified in Figure~\ref{MSC comp}, the preamble accounts for 34.6\% (peaking at 52.9\%) of the little core’s execution time in each check.
Importantly, most of these trap-handling operations are unnecessary in the context of runtime verification, where hardware interactions are disallowed inside checking segments.

While customizing the trap handler is feasible, most of the standard trap-handling sequence must remain to preserve architectural compliance, restricting achievable improvements~\cite{you2011exception, david2006exception, priyadarshan2020exception}.
Moreover, software-level customization requires intrusive modifications to the OS and firmware (infeasible if closed-source):
    (a) adding trap identifiers (\ie, exception code) for cross-privilege check tasks; 
    (b) reconfiguring delegation registers (\eg, \texttt{medeleg} or \texttt{mideleg}) to redirect traps to the correct privilege domain; 
    (c) inserting routines into trap handlers to manipulate PC and branch to the entry of a checking segment; and
    (d) updating the trap vector table when vectored mode is enabled.
Such modifications may span hundreds of lines of low-level code across OS and firmware, forcing persistent maintenance: any update requires revalidation, risking compatibility with existing trap logic and portability across systems.

\begin{tcolorbox}[
  colframe=black,
  boxrule=0.5pt,
  arc=1mm,
  left=1mm,
  right=1mm,
  top=0.5mm,
  bottom=0.5mm,
  boxsep=0.5mm
]

    \textbf{\emph{Insight \#3}:} \textit{
    Utilizing OS-native trap handlers for privileged segment replay incurs excessive preamble overhead and OS invasion.
    Therefore, a hardware-level handler is required to enable low-latency, non-invasive re-execution.}
    
\end{tcolorbox}

\subsection{Resource Dependency Induces Deadlocks}
\label{motiv-deadlock}

Privileged execution often operates on shared OS resources protected by locks, such as memory management (e.g., page tables~\cite{kumar2021pt, gerofi2013pt}) and system timers (e.g., timekeeper~\cite{lamps2014timekeeper, lamps2015timekeeper}). 
This introduces a critical deadlock risk not present in user-only HPED. User-mode and privileged re-execution of big-core work on little cores never accesses memory (or locks), even in multithreaded scenarios, because values are replayed by the LSL. However, replicating user-mode code on little cores can sometimes trigger completely new privileged code that cannot be replayed from the LSL\footnote{Because this is completely new code not run on a big core, it has no redundancy and is not checked for errors. However, this also means it did not influence the execution of the big core's run, and so any errors generated will result in a state mismatch in later checked code if it impacts correct execution.} so must read real data, due to little cores being exposed to some OS abstractions. 

\begin{figure}[t]
     \centering
     \includegraphics[width=1\columnwidth]{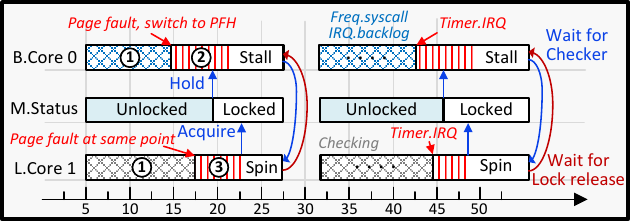}
     \caption{Examples of circular waiting chains.}
     \label{deadlock}
    \end{figure}

In one case, if an instruction faults on the little cores (\eg, because it overtakes the big core or the instruction is paged out before the little core reaches it), the page-fault handler may touch page-table entries locked by the big core. 
In another case, even in checker mode the OS still periodically interrupts the core for synchronization and scheduling~\cite{linux-rcu-stallwarn}. If the timer interrupts on the big and little cores occur in very close temporal proximity (\eg, the big core delays handling its timer due to frequent syscalls), the interrupt handler on the little core may try to access the timer's data structure already locked by the big core. Again we get divergence between big and little cores, meaning logging in the LSL cannot solve resource sharing and locking. Both cause the circular dependencies depicted in Figure~\ref{deadlock}: the big core waits for the finite LSL (held by little cores), while little cores need to access shared resources (locked by big core) to continue trap-handling before releasing the LSL. 
    
This deadlock is intrinsic to the interaction between HPED’s verification mechanism and OS resource management.
Therefore, external intervention is needed to break the dependency chain.
Observing that little core's spin-wait for a lock held by the stalled big core is futile, we break the dependency from the little core side by preempting the spin-lock, switching out the LSL immediately to free space and perform the check to release the big core.

\begin{tcolorbox}[
  colframe=black,
  boxrule=0.5pt,
  arc=1mm,
  left=1mm,
  right=1mm,
  top=0.5mm,
  bottom=0.5mm,
  boxsep=0.5mm
]

    \textbf{\emph{Insight \#4}:} \textit{Extending HPED to privileged modes inevitably introduces deadlock. Active intervention to preempt the little core's futile spin-wait is required for deadlock-free privileged re-execution.}

\end{tcolorbox}

\begin{figure*}[t]
   \centering
   \includegraphics[width=\textwidth]{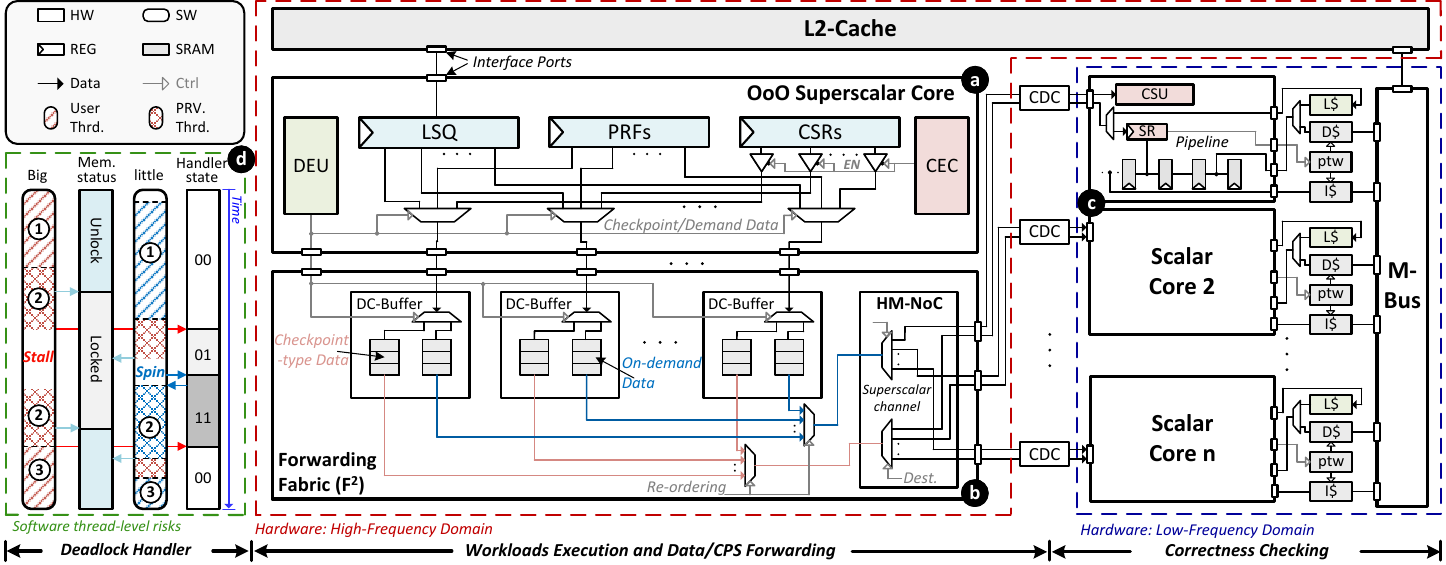}
   \caption{An overview of PEEK (\textit{DEU: Data Extraction Unit; LSQ: Load-Store Queue; PRFs: Physical Register Files; CSRs: Control and Status Registers; CEC: CSRs Extraction Controller; CSU: Context Switching Unit; SR: Shadow Registers; L\$: Load-Store Log}). The data required for program execution and verification is extracted (DEU) and processed (CEC for CSRs) in the big core \blacka{a}, dispatched to the scheduled little core through the forwarding fabric \blackb{b}, and finally replicated and compared in the little cores \blackc{c}. 
   The deadlock handler \blackd{d}  continuously detects and resolves the deadlocks to ensure system liveness.
   }
   \label{overview}
\end{figure*}

%% file: partition/overview.tex
\section{PEEK: At a Glance}
\label{PEEK overview}
Exploiting the bit-level structure and access sparsity of CSRs, we introduce a dynamic snapshot mechanism, which minimizes each RCP's size by eliminating unnecessary bits and CSR synchronization.
On little cores, the privileged checking context switch is completed in hardware, ensuring that the instruction stream bypasses the lengthy trap-handling preamble and directly executes checking segments.
Finally, a generalized solution is developed to handle the cross-core deadlock stemming from contention across diverse shared resources, enabling continuous system operation.



Figure~\ref{overview} presents PEEK's overview, highlighting the key components guided by our motivations.
In the big core (Figure~\ref{overview} \blacka{a}), we deploy a non-invasive Data Extraction Unit~\cite{jiang2025meek} to extract data from the core's Load-Store Queue (LSQ) \textit{on-demand} (i.e. when a load or store occurs so must be logged), \textit{checkpoint-type} data from Physical Register Files (PRFs) (which only changes as a result of repeatable effects, and is collected in checkpoints), and both on-demand and checkpoint-type data from CSRs. CSR extraction is guided by a taxonomy (Table~\ref{tab:csr_classification}), which removes static identifiers and non-repeatable hardware counters from RCPs.
Further, a CSRs Extraction Controller (CEC) is designed, which consists of a static bit-level filter to selectively extract effective CSR bits, and a dynamic access monitor to determine when a CSR should be included in RCPs.
The filter follows the ISA-defined formats of CSRs, where only certain bits are defined while others are reserved or fixed-zero, to only observe bits with architectural information. This further reduces RCP's footprint in bit granularity.
The monitor tracks all CSR accesses from instructions and hardware, with dirty flags reflecting the current status of each CSR. When the latest value of a CSR has been synchronized across all little cores, it is marked as clean and excluded from subsequent RCPs until it is accessed again.
Since some key control registers (\eg, \texttt{pmpcfg} for physical memory access check) may remain clean throughout execution (\ie, setup at boot and never change), that cannot be verified in ECPs, we employ ECC to safeguard the integrity of interaction between these CSRs (Table~\ref{tab:csr_classification}.C4) and corresponding hardware components.

PEEK inherits the Forwarding Fabric (Figure~\ref{overview} \blackb{b}) from MEEK, which employs multiple sets of Dual-Channel Buffers to decouple the delivery of checkpoint-type and on-demand data to the selected little core. This design provides high throughput and stable ordered data transmission even under bursty commit conditions.

In the little cores (Figure~\ref{overview} \blackc{c}), we deploy a set of shadow registers that serves as control and state mirrors for checkpoint-type CSRs (Table~\ref{tab:csr_classification}), replacing instruction interaction of original CSRs while avoiding overhead of context saving and restoring. 
These registers are likewise bit-structured: each mirrors only the effectful bits, minimizing shadow footprint while preserving semantic equivalence.
Moreover, a Context Switching Unit (CSU) is incorporated to assist little cores in context saving and instruction redirection during privileged segment replay. It imitates the standard trap handling to manage the hardware state (e.g., elevating privilege level and disabling interrupt enable before checking). Conceptually, PEEK treats each privileged verification task as a special type of trap, whose entry point corresponds to the starting PC of the checking segment—analogous to \texttt{s/mtvec}. When a privileged segment arrives (\ie, all shadow registers are updated), the CSU directly redirects the instruction to this entry PC and updates associated status bits through hardware gating.
Notably, we disable interrupts in original \texttt{mstatus} but use shadows for instruction interaction (Section~\ref{taxonomy}), making software-level interrupt enabling ineffective—segment replay remains uninterrupted\footnote{The little core re-enables interrupts immediately after completing checking, allowing it to respond to any potentially pending interrupts. While not all privileged segments with disabled interrupts will result in pending interrupts, the latency introduced by a single segment is acceptable for interrupt handling.}, minimizing detection latency. 

Finally, to resolve deadlocks, our solution proactively intervenes on little-core side by detecting and preempting spin-wait through inter-core synchronization mechanisms, and reassigning the little core to perform the verification task the core is stalled on. 
This intervention releases the little core from the wait, allowing it to free the LSL, a key resource required by the big core. Once the LSL is freed, the big core can fill it and proceed to release the system lock, ultimately breaking the circular dependency chain and restoring system-wide forward execution. To this end, we extend the CSU, allowing it to also function as a Deadlock Handler (Figure~\ref{overview} \blackd{d}).


%% file: partition/micro.tex
\section{The Microarchitecture}
Even with the principles in Section~\ref{PEEK overview}, PEEK requires careful microarchitectural engineering to allow precise inter-core synchronization and instruction redirection in little cores with minimal overhead. Privileged code is sensitive, and any state inconsistency or instruction misdirection could lead to catastrophic failures. Therefore, the microarchitecture focuses on guaranteed completeness in state synchronization and correctness in instruction re-execution.
We build PEEK on an open-source heterogeneous SoC (Rocket Chip~\cite{asanovic2016rocket}), featuring both high-performance (BOOM) and energy-efficient (Rocket) cores. We use Rocket Chip as a case study to demonstrate the applicability of PEEK to other heterogeneous SoCs, \eg, ARM's big.LITTLE~\cite{big-little} and Intel's P- and E-class cores~\cite{doweck2017inside}.

\begin{table}[t]
  \centering
  \caption{CSR taxonomy and handling strategy.}
  \label{tab:csr_classification}
  
  \setlength{\aboverulesep}{0.2ex}
  \setlength{\belowrulesep}{0.2ex}

  \begin{tabular}{
    m{0.35cm}
    m{2.7cm}
    >{\raggedright\arraybackslash}m{3cm}
    >{\centering\arraybackslash}m{1.15cm}
  }
    \toprule
    \textbf{Class} & \centering\textbf{Description} & \centering\textbf{Handling} & \centering\textbf{Examples} \tabularnewline
    \midrule
    \textbf{C1} & \textbf{Static identifiers} defining core identity and ISA capabilities. & On-demand extraction. & \makecell{\texttt{mhartid}\\ \texttt{misa}} \\
    \midrule
    \textbf{C2} & Non-repeatable \textbf{Hardware counters} & On-demand extraction. & \makecell{\texttt{cycle}\\ \texttt{instret}} \\
    \midrule
    \textbf{C3} & \textbf{Trap-related} setup \& handling registers. & Add to RCPs, \textbf{replace inst.interaction.} & \makecell{\texttt{mstatus}\\ \texttt{mcause}} \\
    \midrule
    \noalign{\vskip-3pt}
    \textbf{C4} & \textbf{Environmental control} registers. & Add to RCPs, \textbf{full replacement.} & \makecell{\texttt{mcounten}\\ \texttt{menvcfg}} \\
    \bottomrule
  \end{tabular}
\end{table}

\subsection{CSR Taxonomy}
\label{taxonomy}
CSRs differ from architectural registers in that, beyond instruction interaction, they are tasked with control, status tracking, and hardware interaction.
In HPED, RCPs must (i) reconstitute big core's architectural context for execution consistency and (ii) make faults observable for verification.
Not all CSRs meet these criteria, as their roles vary significantly—some are irrelevant to architectural context, while some lack interaction with instruction stream, failing to reflect faults.
Thus, CSRs should be selectively added to RCPs based on their roles. Others can or must be treated like load/store data, which is extracted and forwarded only during instruction interaction to ensure correct instruction replay in little cores.
Furthermore, replicating privileged execution does not require \textbf{\textit{fully replacing}} (\ie, replacing both the original control functionality and instruction interaction) checkpoint-type CSRs in little cores. 
Instead, only key control registers that govern execution permissions are required to establish an equivalent execution environment; the rest can be effectively simulated just based on their interactions with instructions reading and writing them. 
Thus, a systematic CSR taxonomy is necessary to guide the extraction in big core (\ie, checkpoint-type or on-demand) and the deployment in little cores (\ie, replacing fully or instruction interaction only).

As shown in Table~\ref{tab:csr_classification}, the CSRs are categorized into four types based on their roles in context establishment and fault observability.
Class-1 (\textbf{\textit{Static Identifiers}}) defines core identity (\eg, \texttt{mhartid}), which is strictly read-only. 
Since these registers act solely as a static reference for instructions, they are extracted on-demand\footnote{While the writable \texttt{misa} controlling ISA extension greatly affects execution consistency, HPED enforces homogeneous ISA, so that all cores can execute the same instruction stream~\cite{ainsworth2018parallel}. Thus, we classify it as an \textit{identifier}.}.
Class-2 (\textbf{\textit{Hardware Counters}}) includes programmable counters that track hardware events. Since these counters are typically updated at cycle level through distinct events in different cores, which cannot be captured in checkpoint-type snapshots as they are non-repeatable for little cores, they are also extracted on-demand.
Class-3 (\textbf{\textit{Trap-related}}) registers are responsible for the setup and handling of traps. 
They interact most frequently with instructions among the four classes.
On the other hand, for little cores, which enter a trap for verifying the big core's handling rather than responding to a standard hardware event, retaining independent control over these CSRs is essential to correctly execute and return from verification traps -- and their values must be correct at the end of checkpoints to guarantee correct future execution by induction principles.
Therefore, we add them to RCPs for comparison, but during execution only their read/write interactions in instructions are reproduced by little cores to facilitate repetition.
Class-4 (\textbf{\textit{Environmental Control}}) registers establish the permission boundaries (\eg, \texttt{mcounteren} for counter accessibility). 
While they are typically configured at boot and remain stable during execution, any divergence in these settings would cause the same instruction to behave differently on the two cores (\eg, a legal counter read on big core becoming illegal on little cores). 
Hence, we add them to RCPs, fully replacing registers to comprehensively replicate CSR behavior.

\subsection{OoO Superscalar Core (The Big Core)}
\label{big core micro}
\begin{figure}[t]
  \centering
  \includegraphics[width=1\columnwidth]{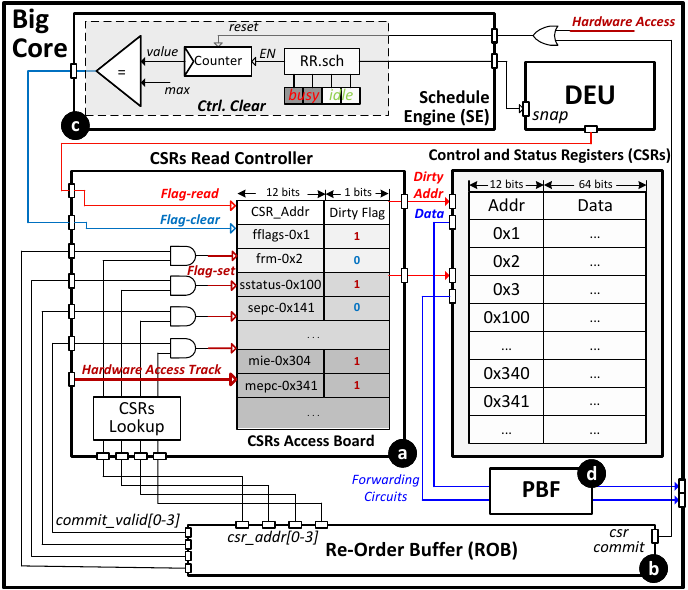}
  \caption{Big core microarchitecture, reducing CSRs redundancy: \blacka{a} records CSR's status in an access board; \blackb{b} tracks instruction retirements to set the board; \blackc{c} generates a clear signal after all little cores are updated; \blackd{d} filters out invalid bits, with parameterization to satisfy specific requirements.}
  \label{big core}
\end{figure}

Program replay requires the collection of both checkpoint-type and on-demand data from big cores. While previous work~\cite{jiang2025meek} has implemented the extraction of checkpoint-type data from PRFs and on-demand data from LSQs, PEEK must extract both types for CSRs while minimizing redundancy. To achieve this, we develop a CSR Extraction Controller (Figure~\ref{big core}), containing a CSR Read controller (CRC) and a Parameterized Bit Filter (PBF). When an RCP is reached, the DEU accesses CRC to selectively read the changed CSR values from corresponding addresses. These values are then filtered and reconstructed by PBF, retaining only relevant bits, and finally delivered to little cores. 
Leveraging CRC's dynamic tracking and the PBF's bit-level extraction, this design greatly reduces CSR volume, mitigating bandwidth consumption.

\noindent
\textbf{Dynamic CSR Read Controller.}
The CRC monitors whether a CSR holds the latest value by maintaining a CSR Access Board (Figure~\ref{big core} \blacka{a}), which tracks the status of each CSR using a dirty bit.  
A set dirty bit indicates that this CSR has been updated by the big core but has not yet been synchronized to all little cores. Board setting is managed by the ROB: we insert an observation channel into the ROB (Figure~\ref{big core} \blackb{b}). When a CSR instruction is retired, its address is routed to a Lookup Table to check for a match with entries in the board. Once a match is found, the corresponding flag is set to 1.
The board is cleared through a control signal generated from the circuit in the Scheduling Engine (Figure~\ref{big core} \blackc{c}). Specifically, we add a counter to the round-robin scheduler\footnote{More precise would be to add a counter to each board entry, to avoid interference from other CSR instructions retiring during scheduling. As seen in Figure~\ref{pattern}, CSR accesses are typically clustered within a segment or adjacent segments rather than evenly distributed. Fine-grained control incurs high logic overhead for negligible benefit.}. 
Each time a new little core is scheduled (\ie, the latest value has been updated to this little core), the counter is incremented. After a scheduling cycle, the counter reaches the number of little cores, then a clear signal is generated. During this process, if any CSR instruction retirement is observed, the counter is reset.
It should be noted that, in addition to CSR instructions, CSRs can be directly modified by hardware. 
For instance, on trap entry the hardware writes \texttt{m/spp} in \texttt{mstatus} with the previous privilege mode, so that m/sret can restore that mode.
Such hardware-induced modifications also set the corresponding board entry and reset the scheduling counter.

\noindent
\textbf{Static Parameterized Bit Filter.}
The PBF (Figure~\ref{big core} \blackd{d}) filters out undefined or fixed-zero bits  for each dirty CSR based on ISA-defined formats. For example, only 33 bits (\eg, \texttt{M/SIE}, \texttt{XS}, \texttt{SUM}, \etc) in \texttt{mstatus} have been defined to keep track of or control the hart’s current operating state, while the remaining 31 bits are reserved \texttt{WPRI} (Writes Preserve, Reads Ignore) fields for future use~\cite{RISCV2025}. This mechanism can also apply to bits that are defined but not fully utilized. A key example is the Exception Code field in the \texttt{mcause} register. Although this field occupies the least significant MXLEN-1 bits (63 bits in RV64), only a small portion is actually used (\eg, just 5 bits for standard exception code, with maximum code 19 indicating the hardware error exception). Therefore, we only need to extract the bits that contain complete and useful information.

To accommodate different field definitions across various processors (\eg, some embedded platforms feature custom interrupt codes), we parameterize the bit filter in hardware. By making it configurable, we enable customization of the extracted fields based on specific demands, allowing the filter to adapt to diverse use-cases, ensuring the extraction of all relevant bits.

\noindent
\textbf{On-demand-type CSRs.}
Our observation channel in the ROB also extracts on-demand-type CSRs, as classified in Table~\ref{tab:csr_classification}. When the channel detects the retirement of these CSRs, forwarding circuits directly read their values from the corresponding locations and transmit them to the on-demand path of the forwarding fabric. 


\subsection{In-order Scalar Core (The Little Core)}
\label{little micro}
\begin{figure}[t]
  \centering
  \includegraphics[width=1\columnwidth]{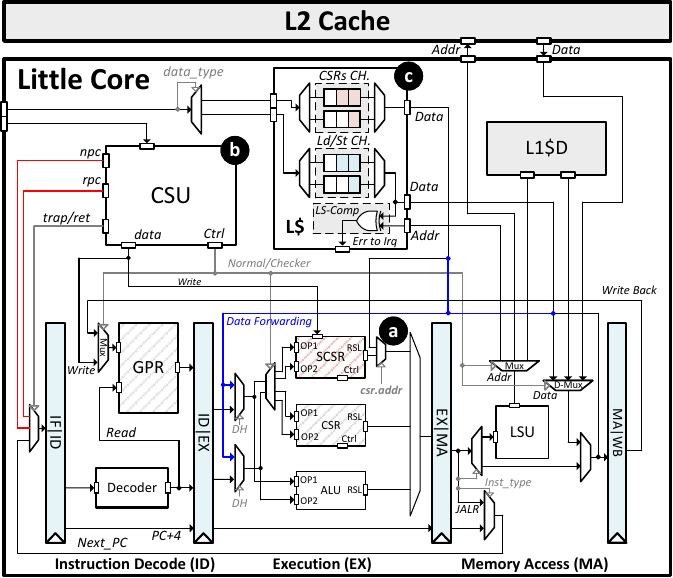}
  \caption{Little core microarchitecture, upgraded with SCSR and CSU: the \blacka{a} SCSR establishes the privileged state mirror of big core; the \blackb{b} CSU serves as the control engine to manage execution mode and instruction redirection. \blackc{c} Decoupling memory and CSR channels eliminates access conflicts.}
  \label{little core}
  \vspace{-3pt}
\end{figure}

Precisely replaying every fragmented privileged checking segment in little cores is crucial to the entire verification pipeline, which requires close hardware coordination of three key operations: privileged state replication, execution mode switching and instruction redirection.
To achieve this, we upgrade the little core's microarchitecture (Figure~\ref{little core}, featuring a 5-stage pipeline) with a set of Shadow CSRs (SCSR, Figure~\ref{little core} \blacka{a}) to maintain a replica of the big core's privileged state, and a Context Switching Unit (CSU, Figure~\ref{little core} \blackb{b}) to manage the mode switching and instruction redirection in each RCP. 
During re-execution, the results of \texttt{Load} and on-demand-type CSR instructions are fetched from the LSL (Figure~\ref{little core} \blackc{c}). As memory and CSR operations typically occur at different pipeline stages, we decouple the Load-Store and CSR channels (\ie, dual-way FIFOs) in the LSL to eliminate potential pipeline stall caused by access conflicts between the two instruction types.

\noindent
\textbf{Shadow CSRs.} 
Since the architectural registers are usually initialized by overwriting, incurring additional costs of context saving and restoring, we use shadow mirroring for CSRs (Figure~\ref{little core} \blacka{a}) to store checkpoint-type data and replace the original CSR according to the strategy detailed in Section~\ref{taxonomy}. This allows us to employ a simple multiplexer that selects between the original and shadow CSRs based on execution mode, enabling rapid privileged context switching. 
While shadow mirroring is theoretically applicable to architectural registers, our synthesis results indicate that the area overhead of copying a complete set of them is enormous, accounting for 15\% of a little core, which is unacceptable for HPED.
Therefore, we retain an SRAM structure for architectural registers, which occupies less than 5\% of the area.
In contrast, thanks to our reduction of CSR redundancy at different granularities (Section~\ref{PEEK overview}), the SCSR achieves a complete mirror with only a 2\% overhead.

Building on the design of our PBF (Section~\ref{big core micro}), the SCSR is also bit-structured, so the data must be reconstructed into standard formats when they interact with software and hardware. Besides, SCSR's storage structure is tightly coupled with PBF to achieve minimal overhead, so we parameterize the SCSR in the same way to satisfy flexible application requirements.

\noindent
\textbf{Context Switching Unit.} 
CSU (Figure~\ref{little core} \blackb{b}) serves as a control engine to manage little core's operating state and perform hardware-assisted context switching.
Specifically, when a privileged SCP arrives and initiates the replay of a privileged segment in a little core, the CSU records and overwrites its architectural register files, and switches the operating mode to checker mode.
In addition, similar to a standard trap response, the CSU switches the little core's privilege mode to match that of the privileged segment, and records current operating state in related CSRs, such as \texttt{s/mstatus} and \texttt{s/mepc}. A key difference is that it redirects the entry point from the trap vector (\eg, \texttt{s/mtvec}) to the starting PC of the segment, bypassing lengthy preambles. 
These operations are completed within the same cycle, ensuring that the little core has necessary permissions to fetch the various privileged instructions without causing an access exception. 
After replay, the CSU restores the context from SRAM and corresponding CSRs, and returns the little core to normal mode.

%% file: partition/deadlock.tex
\section{Deadlock Handling}
Our handling strategy of the privilege-induced deadlock (outlined in Section~\ref{motiv-deadlock}) is achieved by reusing and orchestrating two key hardware modules that exist in PEEK: the Scheduling Engine (SE) in big core for deadlock determination, and the CSU in little cores for spin-wait preemption. 
Specifically, once a circular dependency chain is detected in SE, the CSU drives preemption to redirect little cores from spin-wait to verification tasks, thereby freeing the LSL, unblocking the big core and ultimately restoring normal execution.
The remainder of this section details this workflow.

\noindent
\textbf{Deadlock Detection.}
The deadlock is identified when two conditions coincide: (i) all little cores are stuck in spin-wait, and (ii) the big core is stalled due to unavailable little cores and stops delivering new checking segments to them. Accordingly, we deploy a commit counter at the write-back stage of each little core to track the number of retired lock-checking instruction loops. Once the count exceeds a preset threshold, a status flag indicating the little core's spin-wait state is sent to the SE\footnote{The threshold is configurable and is set to a low value (\eg, 20) to minimize deadlock detection latency without risk of misidentification, as the final decision rests with the big core. That is, a little core's spin-wait does not always signal a deadlock; \eg, the lock is held by another progressing little core, in which case the system can resolve it autonomously. Thus, the big core’s SE, which maintains a global view of all cores, conducts this final decision to eliminate misidentification.}. 
In the big core, a 2-bit FSM (Finite State Machine, Figure~\ref{overview} \blackd{d}) is added to the SE, which is controlled by the status flags from little cores and the scheduling signals. The FSM transitions to state 11 when the SE determines that both conditions are satisfied, which flags a deadlock.

\noindent
\textbf{Preemption \& Context Switching.}
If a deadlock is identified by the big core, a preemption signal is generated from the SE and broadcast to all little cores through the forwarding fabric. 
This signal acts as a non-maskable interrupt that directs the CSU to actively switch the little core's operating state from spin-wait to replaying the pending checking segment (the specific process has been described in Section~\ref{little micro}). Regardless of whether the pending segment is intact and has not yet begun checking, or was previously interrupted by the trap (\eg, a page-fault exception), checking resumes on the recovered little core, ensuring that no instruction can escape the verification. 
When the segment is finished and the LSL is cleared, the little core becomes available, allowing the big core to deliver subsequent segments. The big core can then resume its forward execution, freeing the shared resource, which ultimately resolves the deadlock.
Following each replay of a checking segment, the little cores attempt to re-acquire the lock, ensuring timely detection of whether the big core has released it. 

\begin{table}[t] 
    \centering
    \renewcommand{\arraystretch}{1} 
    \caption{Hardware configurations evaluated.}
    \label{tab:hardware_config}
    
    %
    %
    
     \begin{tabularx}{\linewidth}{
        >{\bfseries\raggedright\arraybackslash}m{1.6cm} 
        >{\raggedright\arraybackslash}X
    }
        \toprule 
        
        \multicolumn{2}{c}{\textbf{Big Core}} \\ 
        \midrule 
        
        \textbf{Core} & { 
                        4-Width, OoO superscalar SonicBoom, @3.2GHz 
                        128-entry ROB, 96-entry IQ, 32-entry LDQ/STQ, 
                        128 Int/FP Phy Registers, 2 ALUs, 1 FPU/Mul/Div
                        } \\
        
        \textbf{Pipeline} & ALU, 2 MEM, 1 Jump Unit, 1 CSR Unit \\
        
        \textbf{Branch Prediction} & \makecell[l]{
                                 TAGE algorithm, 256-entry BTB, 32-entry RAS, \\
                                 6 TAGE table with 2 - 64 bits history
                                 } \\
        
        \midrule 
        
        \multicolumn{2}{c}{\textbf{Memory Hierarchy}} \\ 
        \midrule 
        
        \textbf{L1 ICache} & 32 KB, 4-way, 8 MSHRs , 2 cycle hit lat\\
        \textbf{L1 DCache} & 32 KB, 4-way, 8 MSHRs , 2 cycle hit lat\\
        \textbf{L2 Cache} & 512 KB, 8-way, 12 MSHRs, 12 cycle hit lat\\
        \textbf{Memory} & 16 GB DDR3 @1066MHz, max 32 requests \\
        
        \midrule 
        
        \multicolumn{2}{c}{\textbf{Little Cores}} \\ 
        \midrule 
        
        \textbf{Cores} & \makecell[l]{
                         $4 \times$ In-order Rocket, 5-stage pipeline, @2GHz, \\
                         8-Unroll DIV, 3-stage FPU
                         } \\
        
        \textbf{LSL} & 4 KB, 5000 instruction time-out \\
        \textbf{L1 Cache} & 4 KB, 2-way for both I- and D-Cache \\
        
        \bottomrule 
    \end{tabularx} 
\end{table}

\noindent
\textbf{Post-Resolution Recovery.}
Once the deadlock is resolved by the big core releasing the shared resource, the system undergoes a recovery process to resume normal execution.
This involves two main steps: (i) the FSM in the SE resets to state 00 and ceases generating preemption signals; and (ii) the little cores resume their original trap-handling routines. 
The recovery is triggered when a little core successfully acquires the previously contested lock.
The CSU then restores the architectural context, and the instruction stream is redirected from the \texttt{mepc} to continue the original trap handling. 
Concurrently, the status flag is cleared, prompting the FSM in the big core to reset, which indicates that the system has returned to a deadlock-free state.
Other little cores are similarly restored from the preemptive state and acquire the system resource in turn to complete their respective routines.


%% file: partition/evaluation.tex
\begin{figure*}[t]
  \centering
  \includegraphics[width=\textwidth]{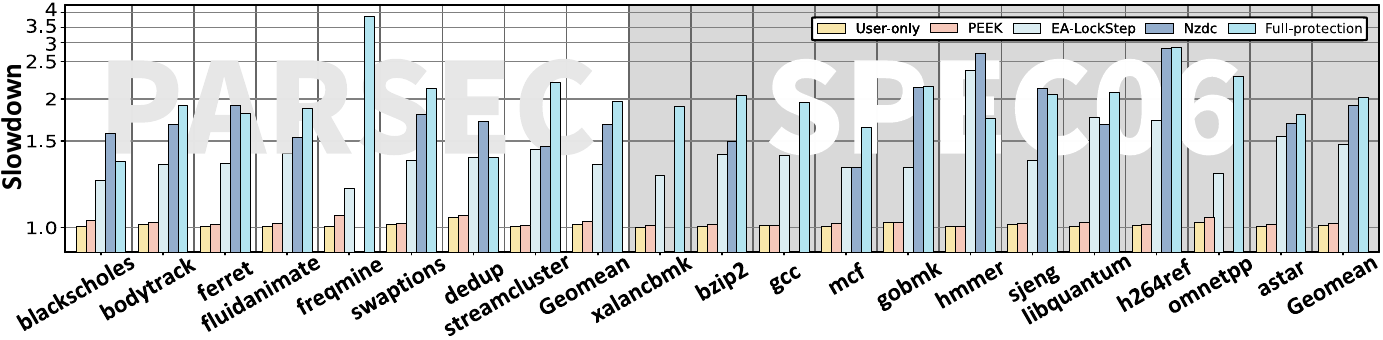}
  \caption{Slowdown for PEEK, EA-LockStep, nZDC~\cite{liao2025deepnzdc}, traditional user- and full-protection~\cite{jiang2025meek}, running SPEC and PARSEC}
  \label{performance}
\end{figure*}

\begin{figure}[t]
  \centering
  \includegraphics[trim=0cm 0.05cm 0cm 0cm, clip,width=1\columnwidth]{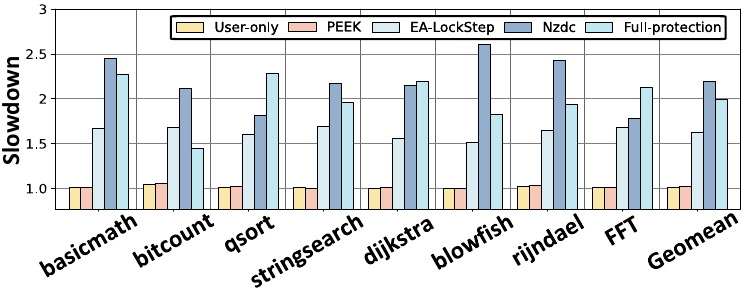}
  \caption{Slowdown when running MiBench.}
  \label{p-mibench}
\end{figure}

\section{Evaluation}
\label{eval}

\subsection{Experimental Setup}
\label{overall setup}

We develop PEEK on an open-source heterogeneous SoC (Rocket Chip~\cite{asanovic2016rocket}), integrating a high-performance  (BOOM~\cite{zhao2020sonicboom}) core and four energy-efficient (Rocket~\cite{asanovic2016rocket}) cores. We implement the microarchitecture with base configurations detailed in Table~\ref{tab:hardware_config} on Chipyard platform~\cite{chipyard} (v1.13.1), and synthesize the RTL using Vivado toolchains (v2023.1). The netlist is deployed on an AMD Alveo U280 FPGA using FireSim~\cite{karandikar2018firesim} and booted the Linux kernel v6.7.

\noindent
\textbf{Baselines.} We select five representative comparators for evaluation: 
the user-only HPED uses the RTL-implementation in MEEK~\cite{jiang2025meek}; the traditional full-protection baseline is achieved by extending MEEK with privileged replay capability, which upgrades its MSU to CSU but snapshots all CSRs, to validate the performance benefits of PEEK; Extended nZDC~\cite{liao2025deepnzdc,didehban2016nzdc,10.5281/zenodo.14678386}, an LLVM-based instruction redundancy scheme, is used as it is the only open-source software scheme; the standard core-level LockStep~\cite{ozer2018error-correlation} with two identical BOOM cores (Table~\ref{tab:hardware_config}), provides full fault coverage with 100\% area overhead and negligible slowdown.
Finally, we construct an Equivalent-Area LockStep (EA-LockStep)~\cite{jiang2025meek} with two downscaled BOOM cores for an iso-area performance comparison. Under the fixed area budget matching PEEK (standard BOOM with four Rockets), we downscale the microarchitectural parameters of BOOM's pipeline, such as decode width,
ROB/LSQ/Issue Queue (IQ) depth. 

\noindent
\textbf{Fault Injection.}
To validate PEEK’s detection capability under fault models that are as realistic as practically possible, we adopt, to the best of our knowledge, the first post-synthesis gate-level fault injection methodology for HPED. We target selected high-impact units for fault injection, including 5 storage structures (\ie, ROB, LSQ, PRF, IQ and rename map table) and 4 execution units (\ie, Integer ALU, Floating-Point ALU, Integer Multiplier/Divider, Floating-Point Multiplier/Divider). As representative core components in pipeline, these target units account for over 80\% of the BOOM core area excluding branch predictors (faults in branch predictors do not cause system failures~\cite{mach2025FJ-nobpd, papadimitriou2023FJ}, since branch mispredictions can be directly repaired in hardware through pipeline flushing) and have been widely investigated in prior fault-injection studies~\cite{chatzopoulos2025FJ, papadimitriou2021FJ, papadimitriou2023FJ, papadimitriou2023-FJ}.
Note that fault injection can in principle be applied to all structures, but such an exhaustive approach is infeasible for an OoO superscalar core, as the combination of fault sites in complex microarchitecture and runtime injection windows leads to an intractable campaign space.  
With our synthesizable RTL implementation, we generate gate-level netlists and inject single-site transient and permanent faults at sampled internal sites (\ie, logic-gate and flip-flop outputs) within the target units.


\begin{figure}[t]
  \centering
  \includegraphics[trim=0cm 0.05cm 0cm 0cm, clip,width=1\columnwidth]{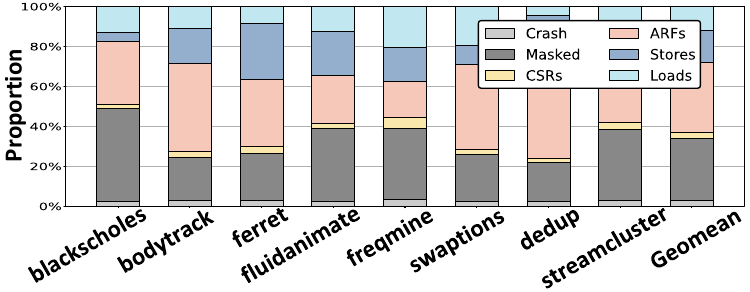}
  \caption{Fault propagation results on PARSEC.}
  \label{fault propagation}
\end{figure}

\subsection{Performance Slowdown}
\label{performance-writing}
\noindent
\textbf{Experimental setup.}
We execute SPECint 2006~\cite{henning2006spec}, PARSEC~\cite{bienia2008parsec} and MiBench~\cite{guthaus2001mibench} on PEEK and compare the performance slowdown versus baseline (results are normalized to standard LockStep). nZDC fails to compile~\cite{liao2025deepnzdc} on some workloads (\eg, gcc).

\noindent
\textbf{Result.}
The slowdown is shown in Figure~\ref{performance} and Figure~\ref{p-mibench}. PEEK incurs a geomean slowdown of only 1.016$\times$ in SPECint, 1.026$\times$ in PARSEC and 1.023$\times$ in MiBench, which is slightly higher than user-only protection (1.009$\times$, 1.016$\times$, 1.011$\times$, respectively) and significantly lower than full protection (2.02$\times$, 1.96$\times$, 1.99$\times$)
\footnote{The massive number of RCPs in large-scale workloads like SPECint and PARSEC leads to prohibitive performance overhead in full-protection scheme using traditional snapshot mechanism, even exceeding that of nZDC.}. 
These results demonstrate the effectiveness of PEEK's privilege-aware design, which extends protection to privileged modes with negligible performance overhead.
In contrast, nZDC introduces substantial geomean slowdown of 1.91$\times$, 1.65$\times$ and 2.45$\times$ due to in-thread redundant checking instructions. EA-LockStep incurs geomean slowdown of 1.47$\times$, 1.31$\times$ and 1.63$\times$, which entails approximately 29$\times$, 11.8$\times$ and 27.4$\times$ more overhead than PEEK.

\subsection{Detection Capability}

\noindent
\textbf{Experimental setup.}
Following the methodology in Section~\ref{overall setup}, we inject 50,000 - 100,000 random faults into the target units of generated gate-level netlist for each workload, and track their propagation throughout the system (detected by PEEK, masked by hardware, or lead to crashes). For faults detected by PEEK, we further measure the detection latency, defined as the interval from fault injection to the moment when divergence is detected in little cores.

\noindent
\textbf{Result.} 
As shown in Figure~\ref{fault propagation}, more than 65\% (geomean) of the injected faults can be exposed through divergence observed at ARFs, CSRs, and load/store\footnote{This does not imply that PEEK covers only 65\% of faults. The remaining faults are either hardware-masked or cause system crashes before being detected, and thus do not silently propagate to the final outputs and manifest as system safety threats.}. 
The contribution of CSR divergence is the smallest, because CSRs exhibit the lowest activity and are less likely to appear in ECPs for comparison, which is also consistent with our earlier empirical characterization of CSR behavior. Around 30\% of the faults are masked by hardware, typically because they occur in invalid entries, in combinational paths of execution units that are not activated by the current instruction, or in registers that are overwritten before being used. Finally, less than 2\% of faults lead to system crashes before being explicitly detected by PEEK. This is acceptable from a safety perspective, because crashes are inherently system-visible and can therefore be properly handled with existing system-level recovery mechanisms~\cite{wang2023crash-is-ok, dixit2021crash-is-ok}. 

\begin{figure}[t]
  \centering
  \includegraphics[width=1\columnwidth]{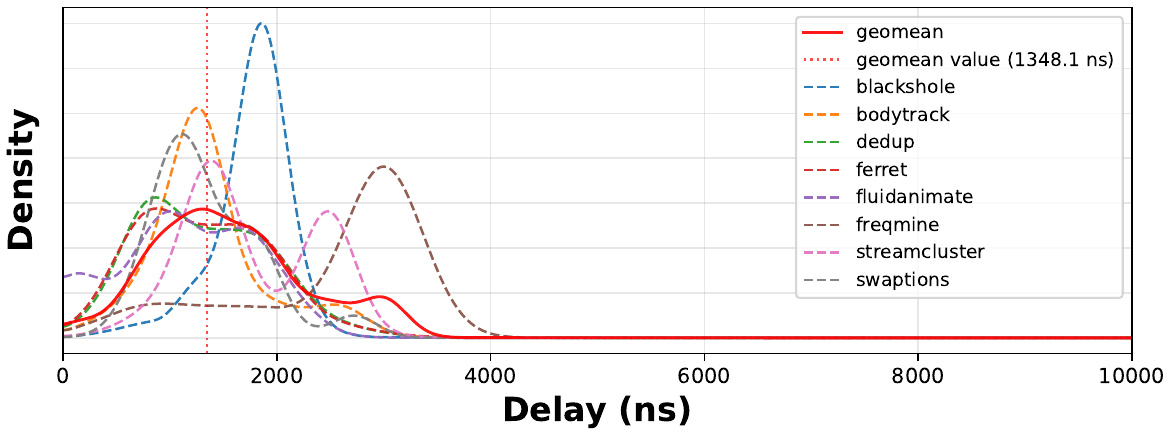}
  \caption{Detection latency while using 4 little cores.}
  \label{latency}
\end{figure}

Figure~\ref{latency} shows the density distributions of the detection latencies across different workloads, each of which features a long, but very thin, tail extending to the far right. 
For most workloads, the majority of detection latencies are concentrated between 1$\mu$s and 2$\mu$s, while the maximum latency seen in freqmine is 2 to 3 times higher, peaking at 4$\mu$s.
Overall, PEEK maintains a detection latency below 4$\mu$s in most cases, which is sufficient to cover over 99\% of hardware faults and multiple orders of magnitude lower than the millisecond-level FTTI requirement for ASIL-D compliance.

\begin{figure}[t]
  \centering
  \includegraphics[trim=0cm 0.05cm 0cm 0cm, clip,width=1\columnwidth]{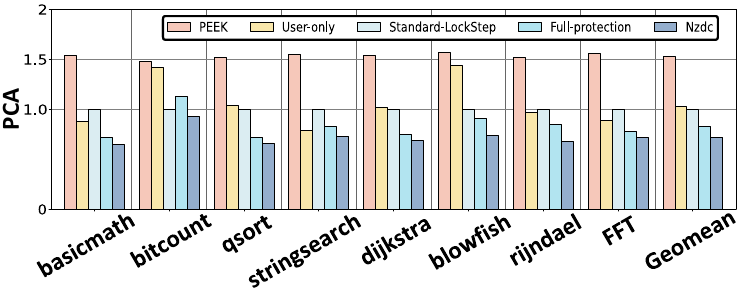}
  \caption{Performance-Coverage-Area metric on MiBench.}
  \label{PCA}
\end{figure}

\subsection{Performance-Coverage-Area (PCA)}
\noindent
\textbf{Experimental setup.} 
To evaluate PEEK's efficiency on multiple key dimensions, we introduce a unified metric, PCA calculated as (IPC $\times$ Coverage $\div$ Area). We compare PEEK’s PCA against our baselines, running privilege-intensive MiBench, as it aligns with our privileged protection focus and fully reflects coverage differences.
EA-LockStep is excluded as it matches PEEK in both area and fault coverage, so a PCA evaluation would only reflect performance differences that have already been evaluated. All PCA values are normalized to the standard LockStep baseline.

\noindent
\textbf{Result.}
As shown in Figure~\ref{PCA}, PEEK's PCA significantly outperforms other comparators in most workloads. Its lead is narrow only in bitcount and blowfish due to their shorter privileged execution times, which limits the coverage benefit of PEEK. nZDC requires the lowest area, yet its high slowdown results in the lowest PCA. These results confirm that PEEK provides the best balance of performance, coverage, and area efficiency among all comparators.

\begin{figure}[t]
  \centering
  \includegraphics[width=1\columnwidth]{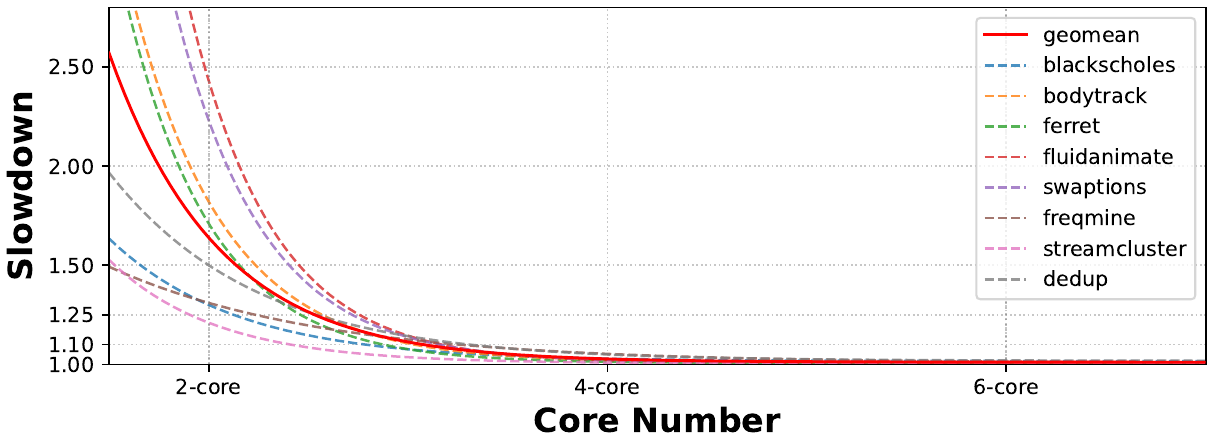}
  \caption{Slowdown under varying numbers of little cores.}
  \label{scal}
\end{figure}

\begin{table}[t]
    \centering 
    \caption{Area overhead comparison.}
    \renewcommand{\arraystretch}{1} 
        
    \begin{tabularx}{\linewidth}{@{}l >{\centering\arraybackslash}X >{\centering\arraybackslash}X >{\centering\arraybackslash}X @{}} 
    \toprule[1pt]

    \textbf{ } & \textbf{User-only} & \textbf{PEEK} & \textbf{Overhead} \\
    \midrule
    Tech. & 28nm & 28nm & \text{---} \\
    Boom & \text{3.89}mm$^2$ & \text{3.96}mm$^2$ & \textbf{\textit{1.8\%}} \\
    Rocket & \text{0.26}mm$^2$ & \text{0.27}mm$^2$ & \textbf{\textit{3.8\%}}  \\
    \midrule
    \textbf{Overhead} & 26.2\% & \cellcolor{green!10}{\textbf{\textit{27.3\%}}}  & \text{---} \\ 

    \bottomrule[1pt]
    \end{tabularx} 
    \label{area}
    
\end{table}

\begin{table}[t] 
    \centering
    \caption{{Power consumption analysis and the breakdown of components designed for privileged protection.
    }}
    \label{tab:power_comparison}
    
    \renewcommand{\arraystretch}{1.0} 
    \setlength{\tabcolsep}{3pt} 
    
    \begin{tabularx}{\linewidth}{>{\raggedright\arraybackslash}X wc{1.8cm} wc{1.7cm} wc{1.6cm}}
        \toprule
        \textbf{{Core (total power)}} & \textbf{{Component}} & \textbf{{Power (mW)}} & \textbf{{Percentage}} \\
        \midrule
        
        \multirow{2}{*}{{\textbf{Rocket (31mW)}}} 
            & {SCSR}                 & {0.66}           & {{\textit{2.13\%}}} \\
            & {CSU}                  & {1.6}          & {{\textit{5.16\%}}} \\
        \midrule
        
        \multirow{2}{*}{{\textbf{BOOM (422mW)}}} 
            & {CEC}                  & {2.2}           & {{\textit{0.52\%}}} \\
            & {DEU}                  & {16.3}          & {{\textit{3.86\%}}} \\
        \midrule
        
        \multicolumn{2}{l}{{{\textbf{Overhead (4*Rocket/BOOM)}}}} & 
        \multicolumn{2}{c}{\cellcolor{green!10}{{\textbf{\textit{29.4\%}}}}} \\
        
        \bottomrule
    \end{tabularx}
    
    \setlength{\tabcolsep}{6pt}
\end{table}

\subsection{Scalability}
\noindent
\textbf{Experimental setup.}
To evaluate scalability, we measure PEEK's slowdown when configured with varying numbers of little cores.

\noindent
\textbf{Result.}
Figure~\ref{scal} shows the slowdown trend against the number of little cores on PARSEC. 
Using two little cores for verification introduces a 1.61$\times$ slowdown (geomean), still outperforming nZDC. 
With four little cores, the geomean slowdown is reduced to 1.026$\times$, with only freqmine and dedup reaching 1.05$\times$.
Six little cores further decrease the geomean slowdown to 1.005$\times$, with all workloads having less than 1.01$\times$.
The observed superlinear trend indicates that PEEK can efficiently scale to more complex workloads and larger big cores by simply incorporating additional little cores. 

\subsection{Hardware Overhead}

\noindent
\textbf{Experimental setup.}
We synthesize the physical implementation of PEEK using TSMC 28nm PDKs and Design Compiler (v2020.12), and employ PrimeTime PX (v2019.12) for power analysis. Our area overhead is contrasted against user-mode-only MEEK~\cite{jiang2025meek}.


\noindent
\textbf{Result.}
As shown in Table~\ref{area}, in PEEK, the BOOM's area is 3.96mm$^2$ and each Rocket is 0.27mm$^2$, introducing 1.8\% and 3.8\% overhead compared to the user-only MEEK, respectively.
From Table~\ref{tab:power_comparison}, integrating SCSR and CSU into Rocket incurs a 7.29\% power overhead, while integrating DEU and CEC into BOOM introduces a 4.38\% overhead.
Versus an unsafe baseline, building PEEK with four Rockets upon a BOOM needs 1.08mm$^2$ and 124mW, \ie, 27.3\% area overhead (slightly higher than MEEK) and 29.4\% power overhead, which is significantly lower than LockStep (\ie, more than 100\% overhead).


\begin{figure}[t]
  \centering
  \includegraphics[width=1\columnwidth]{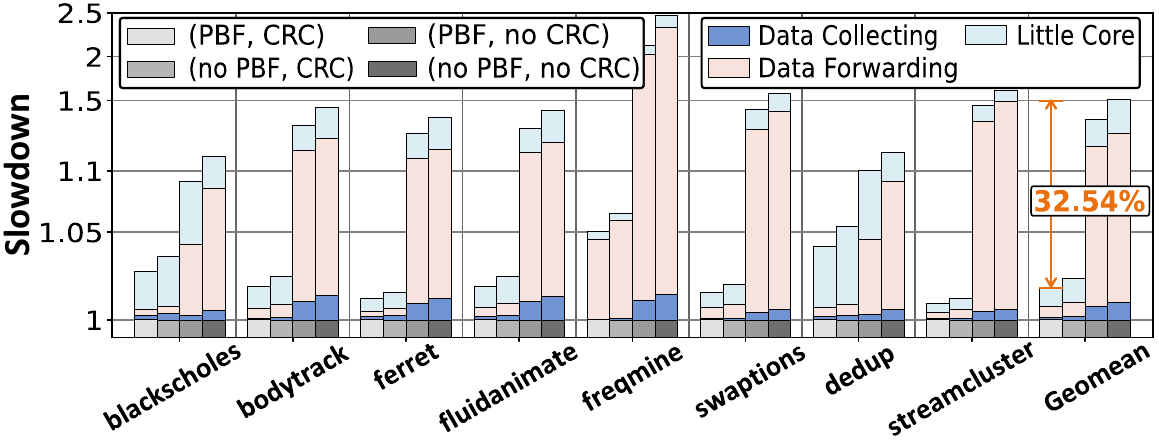}
  \caption{Ablation study and back-pressure breakdown.}
  \label{ablation}
\end{figure}

\subsection{Improvement of CSR-Specific Snapshot}

\noindent
\textbf{Experimental setup.}
To demonstrate the improvement of our proposed snapshot mechanism, we measure slowdown of all four combinations of the PBF and CRC in the big core, which is broken down into three key back-pressure-generating components: data collection, data forwarding and little-core unavailability.  

\noindent
\textbf{Result.} 
As illustrated in Figure~\ref{ablation}, the PBF brings limited performance optimization, because it is static and does not consider the runtime characteristics of CSRs. In contrast, the dynamic CRC introduces a significantly reduced slowdown and shifts the system from being limited by data forwarding to one that is computation-bound, with performance depending on the combined computing power of the little cores. Overall, the PBF and CRC deliver a 32.54\% performance improvement, highlighting that focusing on CSR's unique characteristics is the key to addressing the snapshot bottleneck.

\subsection{Little-Core Utilization}

\noindent
\textbf{Experimental setup.}
To validate the efficiency of our proposed hardware-assisted method for privileged segment replay, we count the verification cycles of the little cores and compare the little-core utilization rate between the OS-native method and ours. To do so, we inject idle cycles equivalent to the execution duration of the preambles (detailed in Section \ref{moti-OS}) into each privileged segment during simulation to simulate replay using the OS-native method. 

\begin{figure}[t]
  \centering
  \includegraphics[width=1\columnwidth]{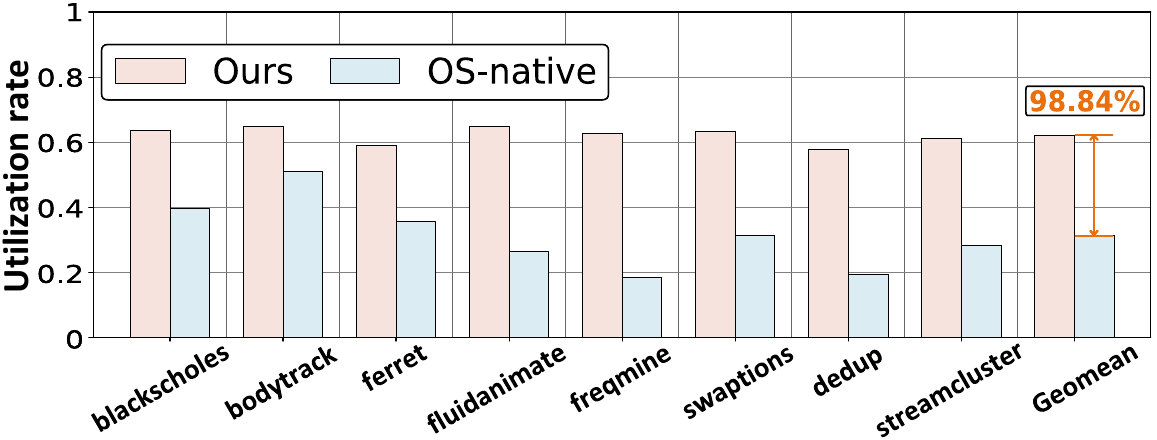}
  \caption{Little-core utilization comparison.}
  \label{utilisation}
\end{figure}

\noindent
\textbf{Results.}
As illustrated in Figure~\ref{utilisation}, our method achieves a utilization of around 60\% in little cores, while the remaining 40\% is used for state transitions, handling their own traps, and idleness (i.e., waiting for verification).
This makes 4 little cores sufficient to keep up with the big core. In contrast, since it spends significant time on preambles, the OS-native method utilizes only about 30\% of little cores, which implies that approximately twice as many little cores (\ie, 8) would be required to match the big core's performance.
 
\subsection{TinyPEEK Die}
\label{tapeout}

Due to die area constraints, we tape out a minimal configuration of PEEK, which we refer to as \textbf{TinyPEEK} with a 2-width BOOM and two Rockets, using SMIC 28nm process. Figure~\ref{Die} (a) shows our chip die, with the TinyPEEK implementation located on the right-hand side. Figure~\ref{Die} (b) shows the area and power breakdown of TinyPEEK.
Due to our significant reduction in configurable components of BOOM, the area and power of the two Rockets account for 52\% and 58\% of BOOM respectively, which are approximately twice those of standard PEEK, but still occupy 48\% less area and consume 42\% less power than LockStep (\ie, more than 100\% overhead).
\begin{center} 
    \includegraphics[width=1\columnwidth]{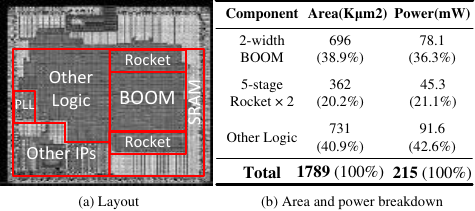}
    \captionsetup{skip=13pt}
    \captionof{figure}{Die with area and power breakdown for TinyPEEK.} 
    \label{Die}
\end{center}

%% file: partition/conclusion.tex
\section{Conclusion}
We have presented PEEK, the first HPED architecture that supports privileged protection in real-world OoO superscalar processors, and also the first to advance HPED from simulation-only research to full-system implementation and silicon tapeout. Our implementation process uncovers a series of challenges in providing privileged protection, including the slowdown explosion caused by full-state synchronization, the privileged segment replay cost, and deadlock. By rethinking the entire verification pipeline through a privilege-aware lens, PEEK achieves full privileged-mode coverage with negligible performance impact, microsecond-level error detection capabilities and minimal hardware overhead. 
The tapeout also solved practical physical design challenges, such as clock domain crossing and timing convergence.
This work not only bridges the gap between HPED theory and real-world deployment but also delivers valuable lessons for both architectural design and physical implementation of future HPED-based reliable systems.